\documentclass{IEEE_lsens}
\usepackage[utf8]{inputenc}
\usepackage{textcomp}
\usepackage[noadjust]{cite}
\ifCLASSINFOpdf
\else

\fi

\usepackage[T1]{fontenc} 
\usepackage{amsmath}

\usepackage[cmintegrals]{newtxmath}

\usepackage{bm}

\usepackage{array}

\usepackage{url}

\usepackage{dblfloatfix}

\usepackage{graphicx}
\usepackage{rotating}
\usepackage{float}
\usepackage{subcaption}
\usepackage{blindtext}
\usepackage{multirow}
\usepackage{graphics}

\ifCLASSINFOpdf

\else

\fi

\providecommand{\hypersetup}[1]{\relax}

\hypersetup{pdftitle={Bare Demo of IEEE\_lsens.cls for IEEE Sensors Letters},
pdfsubject={Typesetting},
pdfauthor={Michael D. Shell},
pdfkeywords={Class, IEEE, IEEE\_lsens, IEEE Sensors Letters, LaTeX, Typesetting, TeX}}

\newcommand{\mat}[1]{\mathbf{#1}}

\begin{document}

\IEEELSENSarticlesubject{}

\title{Effective Acoustic Energy Sensing Exploitation for Target Sources Localization in Urban Acoustic Scenes}

\author{\IEEEauthorblockN{M. Alves\IEEEauthorieeemembermark{1}, R. Coelho\IEEEauthorieeemembermark{2}
and~E. Dranka}
\IEEEauthorblockA{Laboratory of Acoustic Signal Processing, Military
Institute of Engineering, Rio de Janeiro, 22290-270, Brazil\\
\IEEEauthorieeemembermark{1}Student Member, IEEE\\
\IEEEauthorieeemembermark{2}Senior Member, IEEE}\vspace{-0.2cm}}%

\IEEEtitleabstractindextext{%
\begin{abstract}
This letter proposes a new approach to improve the accuracy of the Energy-based source localization methods in urban acoustic scenes. The proposed acoustic energy sensing flow estimation (ESFE) uses the sensors signal nonstationarity degree to determine the area with highest energy concentration in the scenes. The ESFE is applied to different acoustic scenes and yields to source localization accuracy improvement with computational complexity reduction. 
The experiments results show that the proposed scheme leads to
significant improvement in source localization accuracy.
\vspace{-0.7cm}
\end{abstract}

\begin{IEEEkeywords}

Energy-Based Source Localization, Acoustic Scenes, Wireless Acoustic Sensor Network, Index of Nonstationarity.
\end{IEEEkeywords}}

\maketitle

\section{Introduction}
\vspace{-0.25cm}

Wireless Acoustic Source Network (WASN) is a very attractive solution for source localization in urban areas \cite{meng2017energy,lu2014novel}. WASN enables low cost and low power coverage for indoor and outdoor acoustic scene environments. Accurate source location estimation is an ubiquitous issue in a diversity of applications including objects monitoring, seismic event detection, house surveillance and smart vehicle tracking.

Acoustic source localization methods are mainly based on the computation of the time-delay estimation (TDE) or the time-delay of arrival (TDOA) and the acoustic signal energy \cite{cobos2017survey,dranka2015robust,sheng2005maximum,lu2014novel}. TDE or TDOA algorithms use time-delay or phase difference measures obtained at the acoustic sensors generally distributed in a microphone array. Energy-based techniques are simple and interesting solutions widely applied for sound source estimation localization in WASN. The main challenge is the background acoustic interference that can severely affect a target location estimation, particularly when considering real acoustic scenes. Generally, each acoustic scene is composed of multiple sources with different temporal and spectral statistics.

The main goal of this letter is twofold. Firstly, it applies energy-based source localization methods in acoustic scenes environment. And secondly, it introduces an efficient acoustic energy sensing flow exploitation (ESFE) approach for energy-based source localization accuracy improvement. The proposed scheme defines the nonstationary acoustic energy flow formed by individuals sources that composes a scene. The selection of sensors is based on the nonstationarity degree of the collected acoustic sensors amplitude signals. The ESFE enables location estimation accuracy improvement of the Energy-based localization methods with reduced number of sensors. The Cramér-Rao Lower Bound (CRLB) is also derived to examine the robustness of the H-ML-Energy method.

Extensive experiments are conducted to evaluate the effectiveness of the proposed ESFE solution. For this purpose, outdoor Park and indoor Kitchen scenes are simulated using a diversity of real acoustic sources signals. Index of nonstationarity (INS) \cite{borgnat2010testing} of the sensor signals are applied for the sensor node selection in each scenario. The Maximum Likelihood (ML) Energy-based source localization methods, ML-Energy \cite{sheng2005maximum} and H-ML-Energy \cite{dranka2015robust}, are examined before and after the application of the proposed approach. This solution is also compared to a sensor selection method based on noise reduction \cite{deng2017,cobos2017survey}. Experiments are conducted with four different values of SNR (signal-to-noise ratio) ranging from 0dB to 15dB. Experimental results demonstrate that the proposed ESFE scheme improves the accuracy of the energy-based source localization methods while reducing the number of sensors in acoustic scenes with real sources.

\vspace{-0.20cm}
\section{Source Localization in Acoustic Scenes}
\vspace{-0.15cm}

Acoustic scenes are composed of multiple sound sources that naturally belongs to this environment, 
including animals, people and objects.
In order to evaluate the ESFE method, two acoustic scenes\footnote{\label{note1}Available at lasp.ime.eb.br.} were artificially composed of six distinct real omnidirectional acoustic sources randomly placed in the delimited area of each scene. First scene is outdoor Park, that is composed of single sources "speaker", "waterfall", "birds", "dogs barking", "children playing", and "babble". And other scene is indoor Kitchen with sources "speaker", "television", "water", "sizzling", "cutting", and "clanking dishes".

\vspace{-0.35cm}
\subsection{Index of Non-Stationarity}
\vspace{-0.1cm}

\begin{figure*}[!h]
	 \vspace{-0.3cm}
 	     	\centering
	    \begin{subfigure}[b]{0.19\textwidth} 
	        \includegraphics[width=3.5cm]{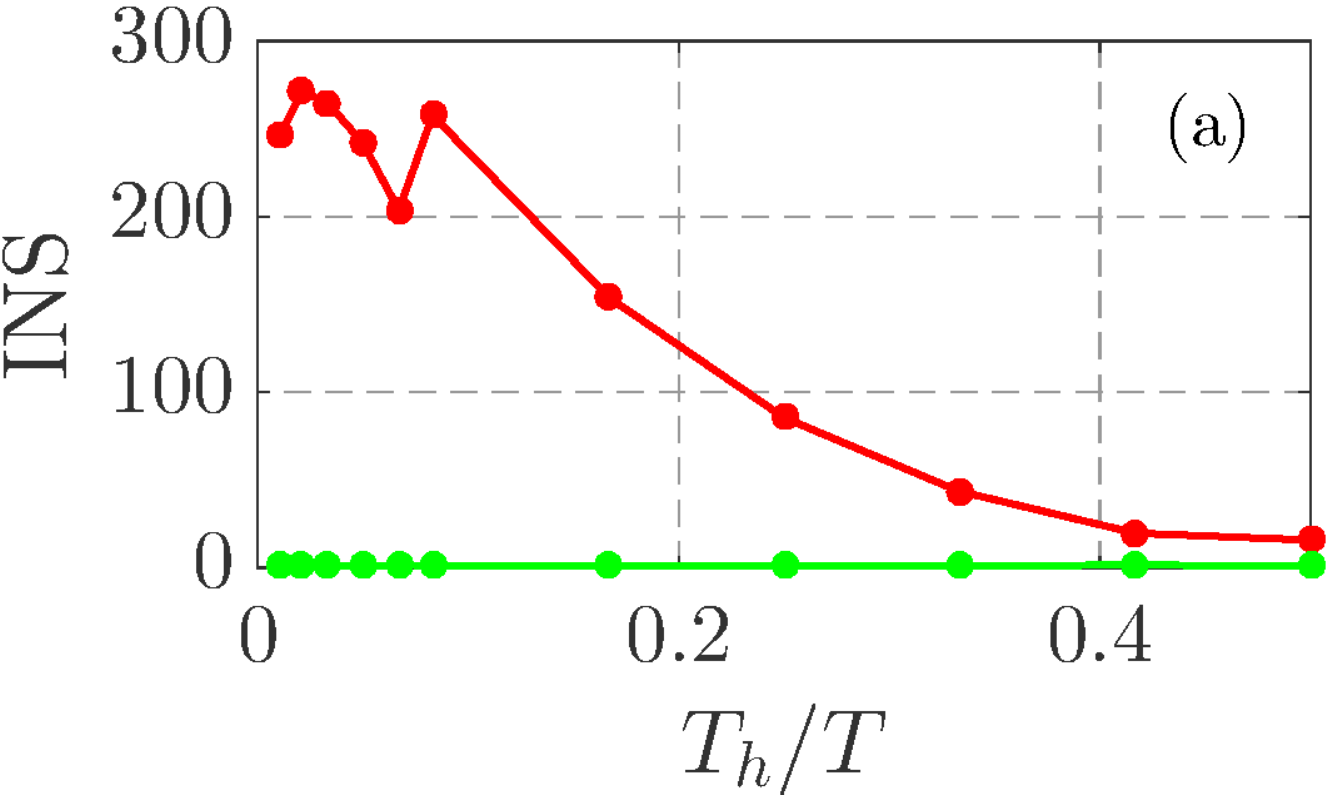}
	        \vspace{-0.5cm}
	     \end{subfigure}	     
	     \begin{subfigure}[b]{0.19\textwidth}  	
	        \includegraphics[width=3.5cm]{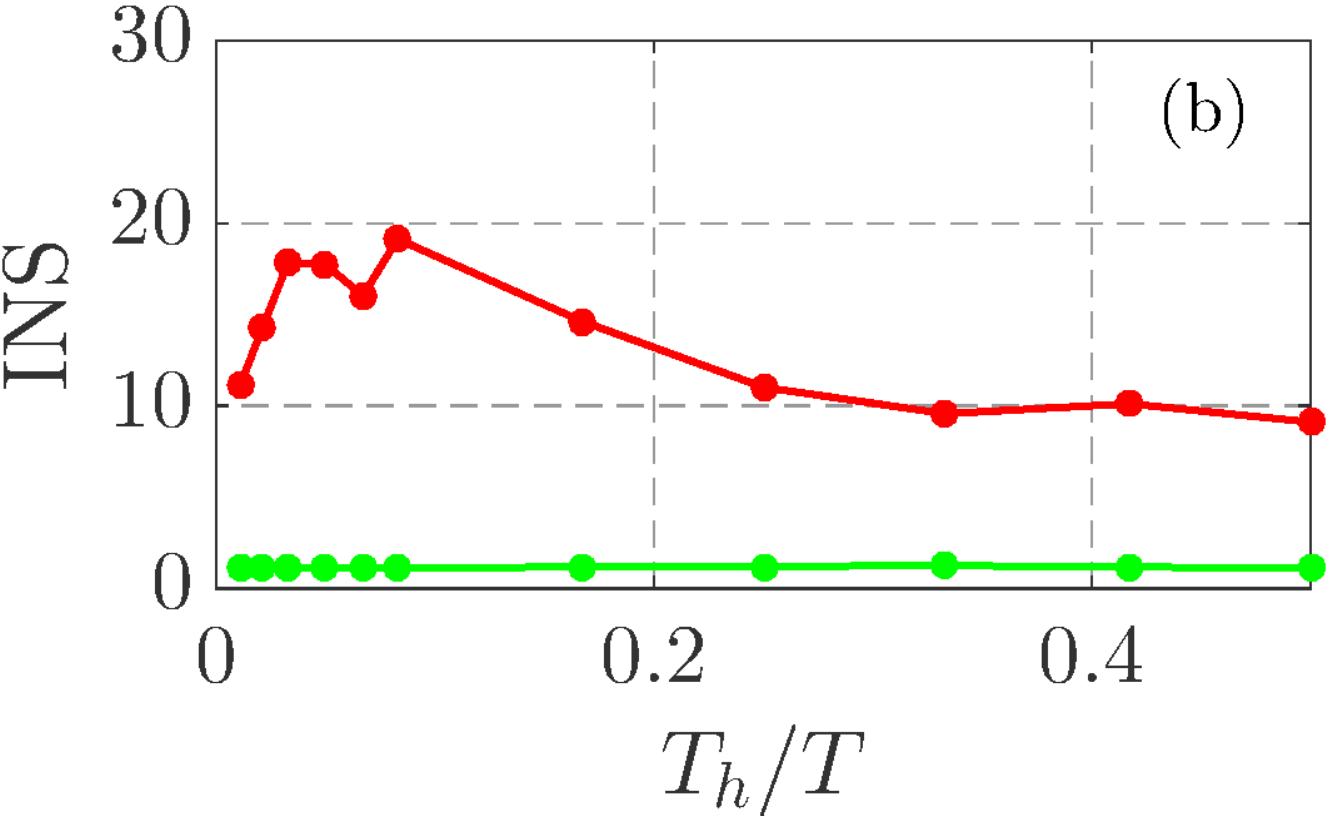}
	        \vspace{-0.5cm}
	     \end{subfigure}	     
	     \begin{subfigure}[b]{0.19\textwidth}
	        \includegraphics[width=3.5cm]{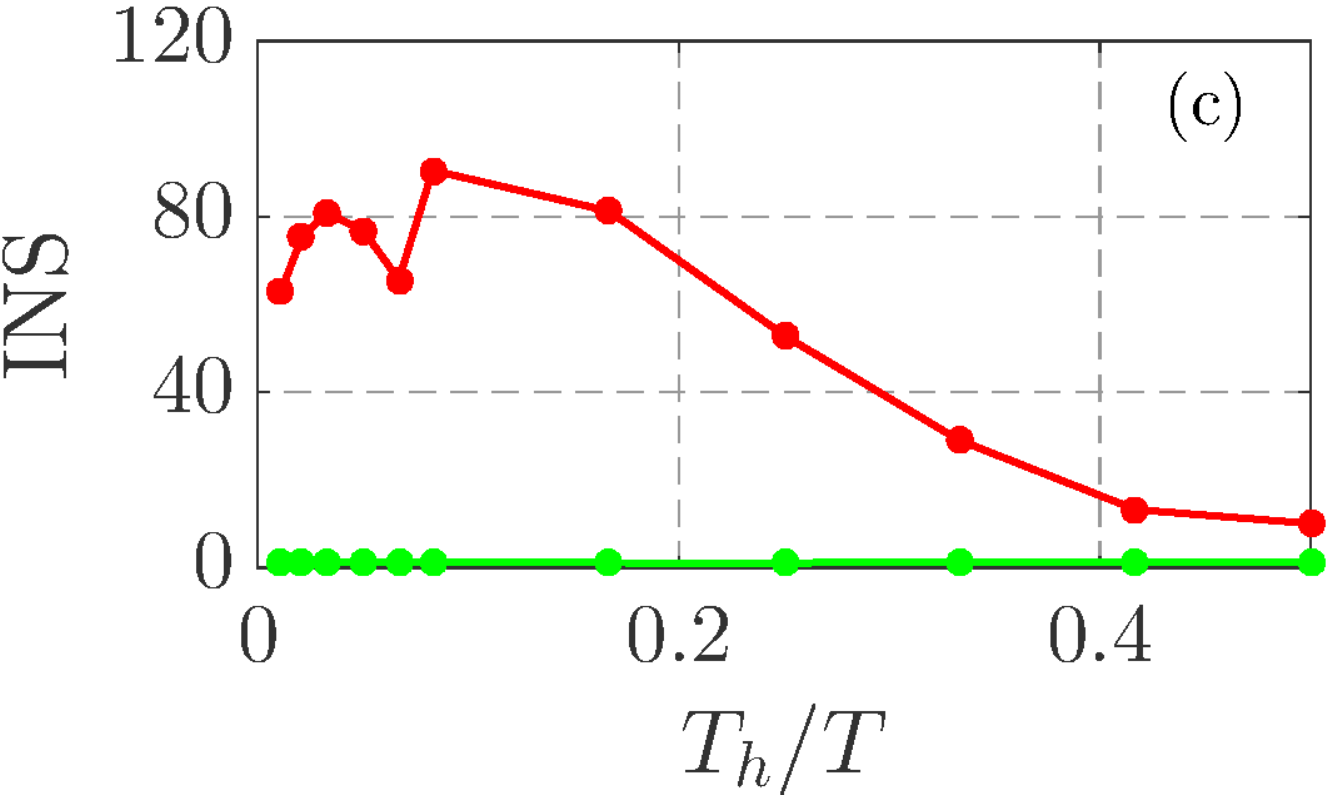}
	        \vspace{-0.5cm}
	     \end{subfigure}
	     \begin{subfigure}[b]{0.19\textwidth}  
	        \includegraphics[width=3.5cm]{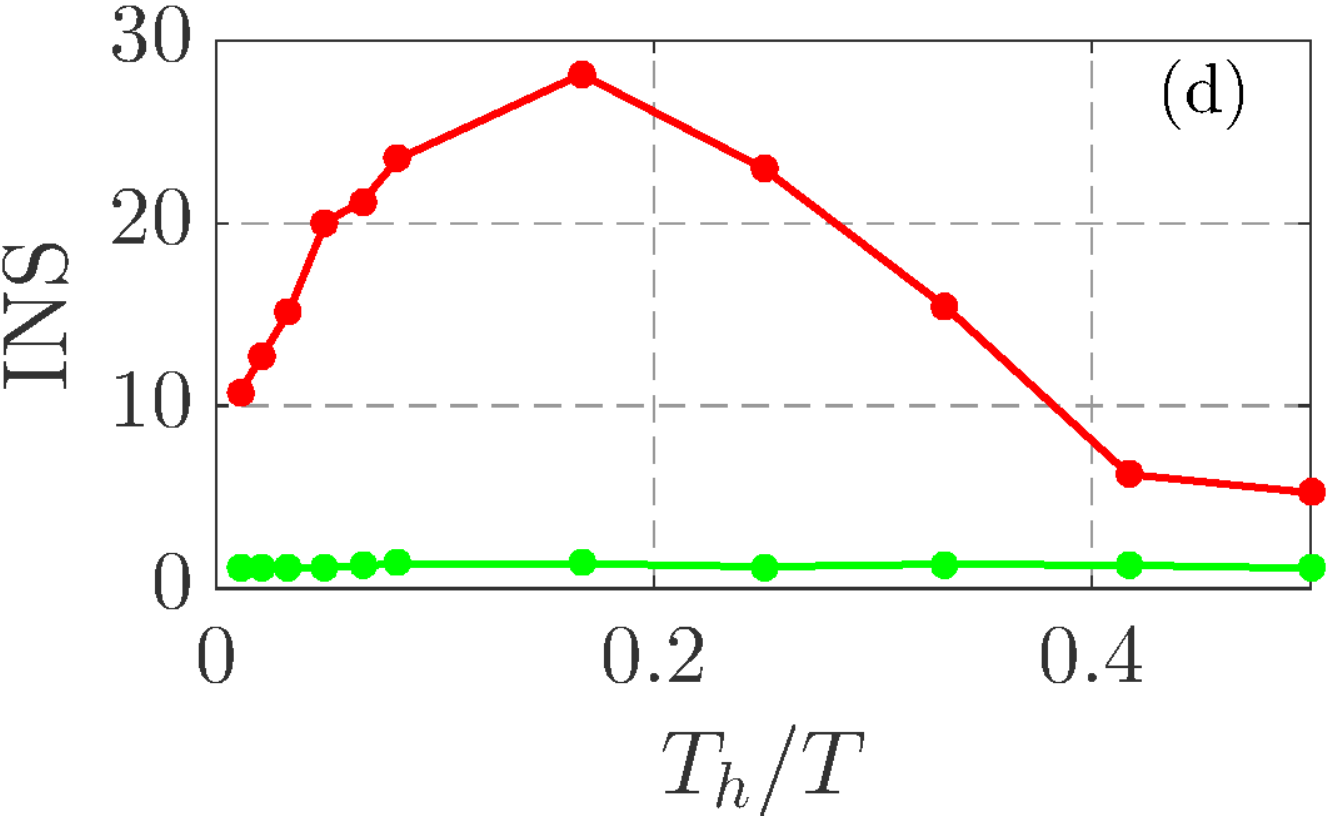}
	        \vspace{-0.5cm} 
	     \end{subfigure}	     
	     \begin{subfigure}[b]{0.19\textwidth} 
	        \includegraphics[width=3.5cm]{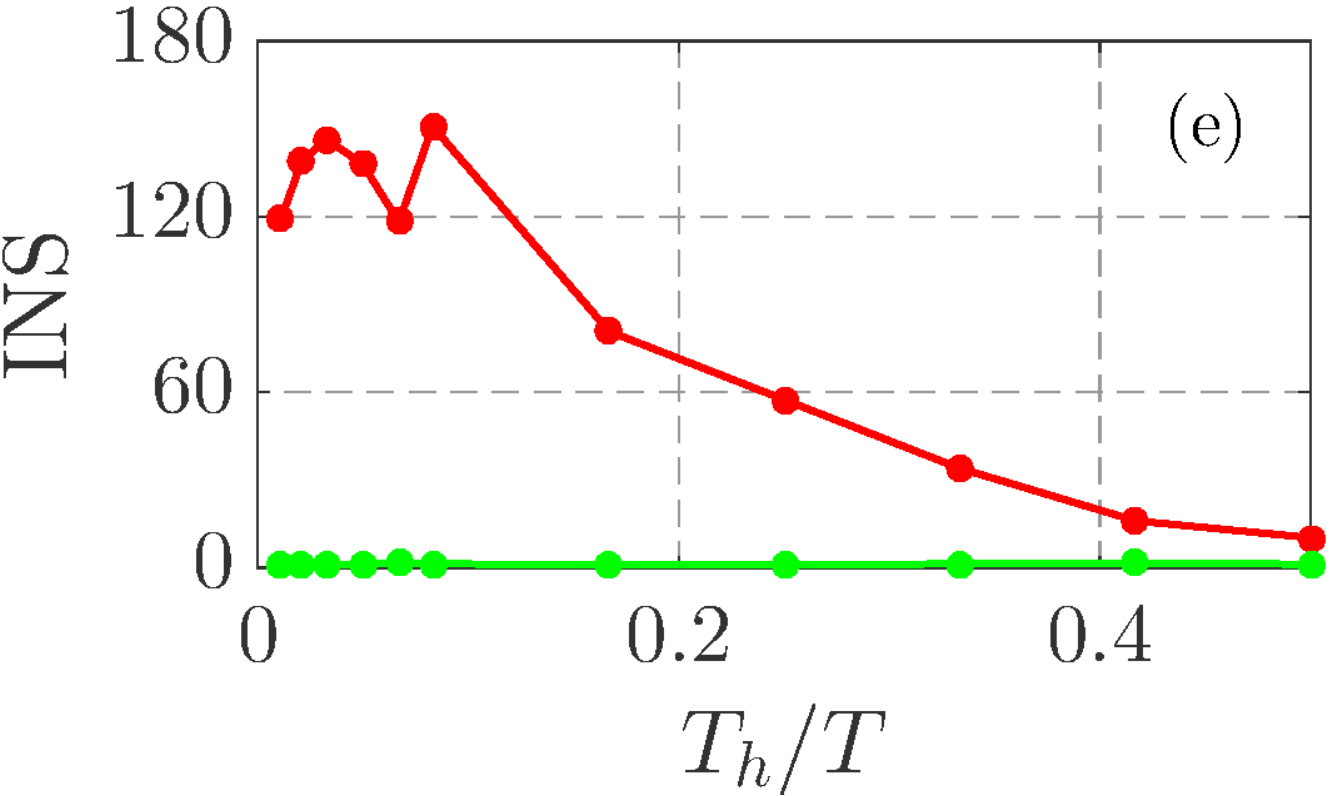}
	        \vspace{-0.5cm}
	     \end{subfigure}
	     \vspace{-0.1cm} 
	     \caption{INS values obtained for 3-seconds audio signals. The red line represents the INS value of each time scale $T_h/T$. The threshold is indicated by the green lines.(a)"speaker", (b)Park, (c)"speaker"+Park (15dB), (d)Kitchen, (e)"speaker"+Kitchen (15dB).}
	     \label{fig:ins}
	    \vspace{-0.3cm} 
	   \end{figure*}
	   
A signal sample sequence is defined as stationary if its main statics are time-invariant. The index of non-stationarity (INS)$^1$ is a time-frequency approach to objectively examine the non-stationarity of a signal. The stationarity test is conducted by comparing spectral components of the signal to a set of stationary references, called surrogates. For this purpose, spectrograms of the signal and surrogates are obtained by means of the short-time Fourier transform (STFT) considering a window length $T_h$. Then, the Kullback-Leibler (KL) divergence is used to measure the distance between the short-time spectra of the analysed signal and its global spectrum averaged over time. Finally, the INS is given by the ratio between this distance and the corresponding KL values obtained from the stationary surrogates. In \cite{borgnat2010testing}, the authors considered that the distribution of the KL values can be approximated by a Gamma distribution. Therefore, for each window length, a threshold $\gamma$ is defined for the stationarity test considering a confidence degree of 95\%. Thus,
\vspace{-0.25cm}
\begin{equation}
\vspace{-0.15cm}
\text{INS}\begin{cases}
    \leq \gamma, & \text{signal is stationary};\\
    > \gamma, & \text{signal is nonstationary  }.
  \end{cases}
\end{equation}

Fig. \ref{fig:ins} illustrates the INS values of five studied signals:
the source "speaker", acoustic scenes Park and Kitchen, and the source corrupted by the scenes as background noise. 
The time scale $T_h/T$ indicates the relation between the length adopted in the short-time spectral analysis ($T_h$) and the total length ($T = 3$ seconds) of the signal. 
Red lines represent INS values and green lines indicate threshold values.
A signal is stationary if its INS value is below the threshold for every time scale. 
Signals with INS values above the threshold for the majority of the time scales are classified as non-stationary.
A signal is considered as highly non-stationary if the maximum INS value, $\text{INS}_\text{max}$, is greater than 100.
The "speaker" source is here classified as highly nonstationary, while the two scenes are nonstationary. 
When corrupted by noise, the $\text{INS}_\text{max}$ of a signal decreases since it becomes more stationary. Note that the "speaker"
source has $\text{INS}_\text{max} \approx 300$. However, its $\text{INS}_\text{max}$ is reduced to 80 and 150 when corrupted by the Park and Kitchen scenes, respectively. In other words, the INS is a representative parameter to indicate if a target signal is corrupted by noise. The ESFE method exploits this INS characteristic for the selection of most representative sensors in the WASN.

\vspace{-0.3cm}
\section{Energy-Based Source Localization in Acoustic Scenes}
\vspace{-0.2cm}
	  
Energy-based source localization methods are based on the fact that the acoustic energy attenuation is inversely proportional to the distance from the signal to the multiple acoustic sensors distributed in the field \cite{kinsler82}. A maximum likelihood approach was presented in \cite{sheng2005maximum} for acoustic source position estimation (ML-Energy), and later in \cite{dranka2015robust} for noisy correlated environments (H-ML-Energy). In this work, the methods ML-Energy and H-ML-Energy are going to be adjusted for the acoustic scene environment.

Localization methods in acoustic scenes are applied to estimate the target source position, 
while the summation of the other sources is considered as noise. The signal received at
the $i$-th sensor is sampled during the $n$-th time interval with a sampling frequency $f_s$. is defined as $x_i(n)=A_i(n)+W_i(n)$, where $A_i(n)$ is the acoustic signal intensity or energy given by
\vspace{-0.2cm}
\begin{equation}
\vspace{-0.2cm}
A_i(n)=\sqrt{g_i}\sum^{K}_{j=1}\frac{s_j(n-\tau_{ji})}{|{\bf p}_j(n-\tau_{ji}) - {\bf r}_i|},
\label{eq:ui}
\end{equation}where $g_i$ represents the sensor gain, the $s_j$ is the signal intensity of the $j-$th source, ${\bf r}_i$ is the sensor position, 
${\bf p}_j$ is the $j$-th ($j = 1, \ldots, K$) source spatial coordinates. In acoustic scenes, the total noise intensity, $W_i(n)$, is defined as
\vspace{-0.3cm}
\begin{equation}
\vspace{-0.2cm}
W_i(n)=\sqrt{g_i}\sum^{M}_{m=1}\frac{o_m(n-\tau_{ji})}{|{\bf w}_m(n-\tau_{ji}) - {\bf r}_i|},
\end{equation}where $o_m$ represents the $m-$th ($m=1,\cdots, M$) noise source intensity and ${\bf w}_m$ its position. Given the time index $t$, the acoustic energy in the $i$-th sensor, $\mathbb{E}[x_i^2(n)]=u_i(t)$, is given by
\vspace{-0.2cm}
\begin{equation}
\vspace{-0.2cm}
u_i(t)=g_i\sum^{K}_{j=1}\frac{B_j(t)}{d^2_{ij}(t)}+2\mathbb{E}[A_i(t)w_i(t)]+\mathbb{E}[W_i^2(t)],
\label{eq:ui2}
\end{equation}where $B_j$ is the $j$-th source acoustic energy, and $d_{ij}$ the distance between the source $j$ and sensor $i$. In the ML-Energy, the background noise is modelled as uncorrelated with the source signal, the cross term $\mathbb{E}[A_i(t)W_i(t)]$ is considered equal to zero. This assumption can severely degrade the sensor measurements and also, the source localization estimation accuracy. 
In \cite{dranka2015robust}, the correlation between the source and the signal is taken in consideration. In the H-ML-Energy method, the cross term $\mathbb{E}|A_i(t)W_i(t)|$ and the term $\mathbb{E}|W_i^2(t)|$ are modelled by a fractional Gaussian noise (fGn) with exponent $H$, mean $\mu_H$ and variance $\sigma_H$, obtained from the sensor readings. 
Denoting the fGn process by $h_i(t)=2\mathbb{E}[A_i(t)W_i(t)] + \mathbb{E}[W_i^2(t)]$, 
the acoustic source localization energy is defined as
\vspace{-0.2cm}
\begin{equation}
\vspace{-0.2cm}
u_i(t)=g_i\sum^{K}_{j=1}\frac{B_j(t)}{d^2_{ij}(t)}+h_i(t).
\end{equation}

The fGn process represents the energy measurement error. Since the fGn is able to indicate any degree of correlation by means of its exponent, the H-ML-Energy grants better accuracy to the energy based localization model. The vector $\mat{Z_H}$ represents the normalized acoustic energy in each sensor $i (i=1, \cdots , L)$. Thus, 
$\mat{Z_H}= [\frac{u_1-\mu_{H_1}}{\sigma_{H_1}} \cdots \frac{u_L-\mu_{H_L}}{\sigma_{H_L}}]^T$. 
The joint probability density function $\mat{Z_H}$ in matrix form is
\vspace{-0.2cm}
\begin{equation}
\vspace{-0.2cm}
    f(\mat{Z_H}|\boldsymbol{\theta})=(2\pi)^{-L/2}\exp \{-\frac{1}{2}(\mat{Z_H}-\mat{G_HDB})^T(\mat{Z_H}-\mat{G_HDB})\},
\label{eq:ml}
\end{equation}where $\mat{G_H}$ represents the gain matrix, $\mat{D}$ is the attenuation matrix, $\mat{B}$ is the acoustic energy source vector 
and 
$\boldsymbol{\theta}= [\mat{\rho}^T_1 \mat{\rho}^T_ 2 \cdots \mat{\rho}^T_K B_1 B_2 \cdots B_K]^T$ is a vector with the source positions $\mat{\rho}_j$ and their corresponding acoustic energies $B_j$. These matrices are defined in \cite{dranka2015robust}.
In this work, multiresolution search \cite{sheng2005maximum} is applied to obtain the minimum value of the log-likelihood function,
\vspace{-0.3cm}
\begin{equation}
\vspace{-0.3cm}
    L(\boldsymbol{\theta})=||\mat{Z_H}-\mat{G_HDB}||^2.
\end{equation} 

	 \begin{figure}[!t]
	 \vspace{-0.1cm}
	    \begin{subfigure}[b]{0.22\textwidth} 
	        \includegraphics[scale=0.08]{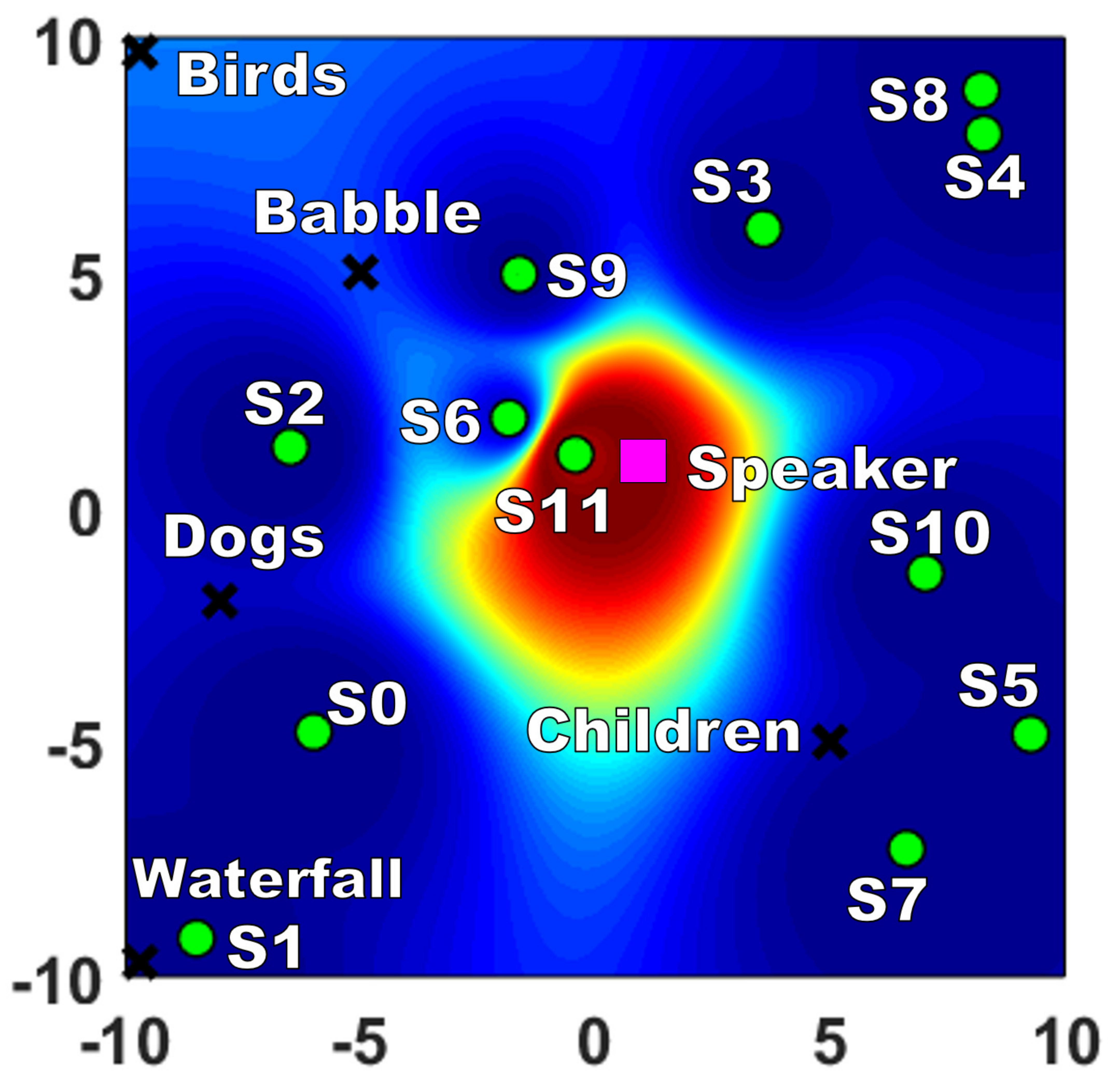}
	        \vspace{-0.2cm}
	        \caption{Park (20m x 20m)}
	     \end{subfigure}
	     \begin{subfigure}[b]{0.24\textwidth} 
	        \includegraphics[scale=0.232]{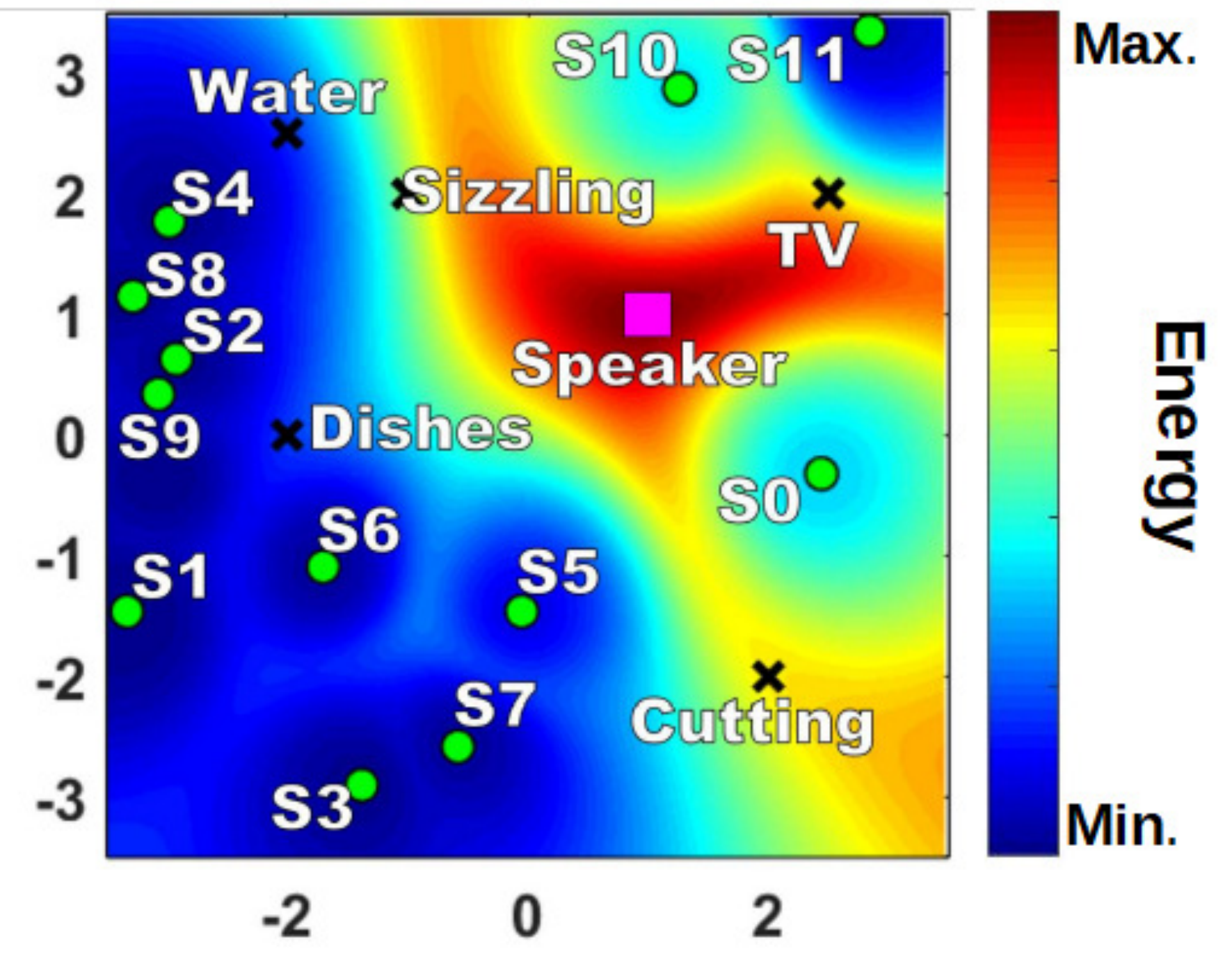}
	        \vspace{-0.2cm}
	        \caption{Kitchen (7m x 7m)}
	     \end{subfigure}
	     \vspace{-0.2cm}
	    \caption{Energy distribution scene maps. Sensors are represented by circles, target source by square, and other/noise sources by "$\times$".}
	    \label{fig:scene}
	    \vspace{-0.4cm}
	   \end{figure}

	   \begin{figure*}[!t]
	 \vspace{-0.6cm}
	     \begin{subfigure}[b]{0.24\textwidth}	         \includegraphics[width=\textwidth, height=3 cm ]{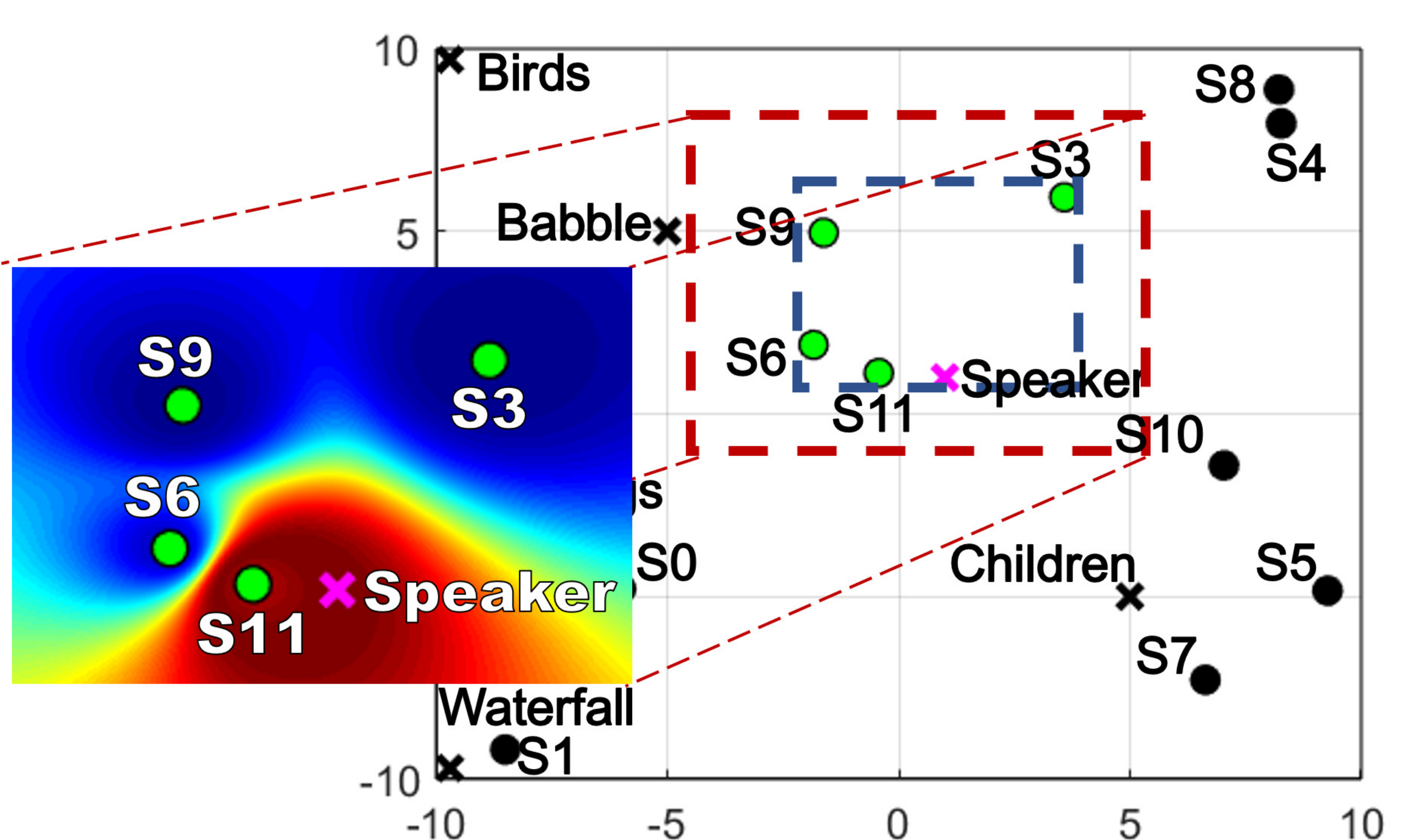}
	     \vspace{-0.5cm}
         \caption{Scene:Park/Source:"speaker"}
         \end{subfigure}
         \begin{subfigure}[b]{0.24\textwidth}	         \includegraphics[width=\textwidth, height=3 cm]{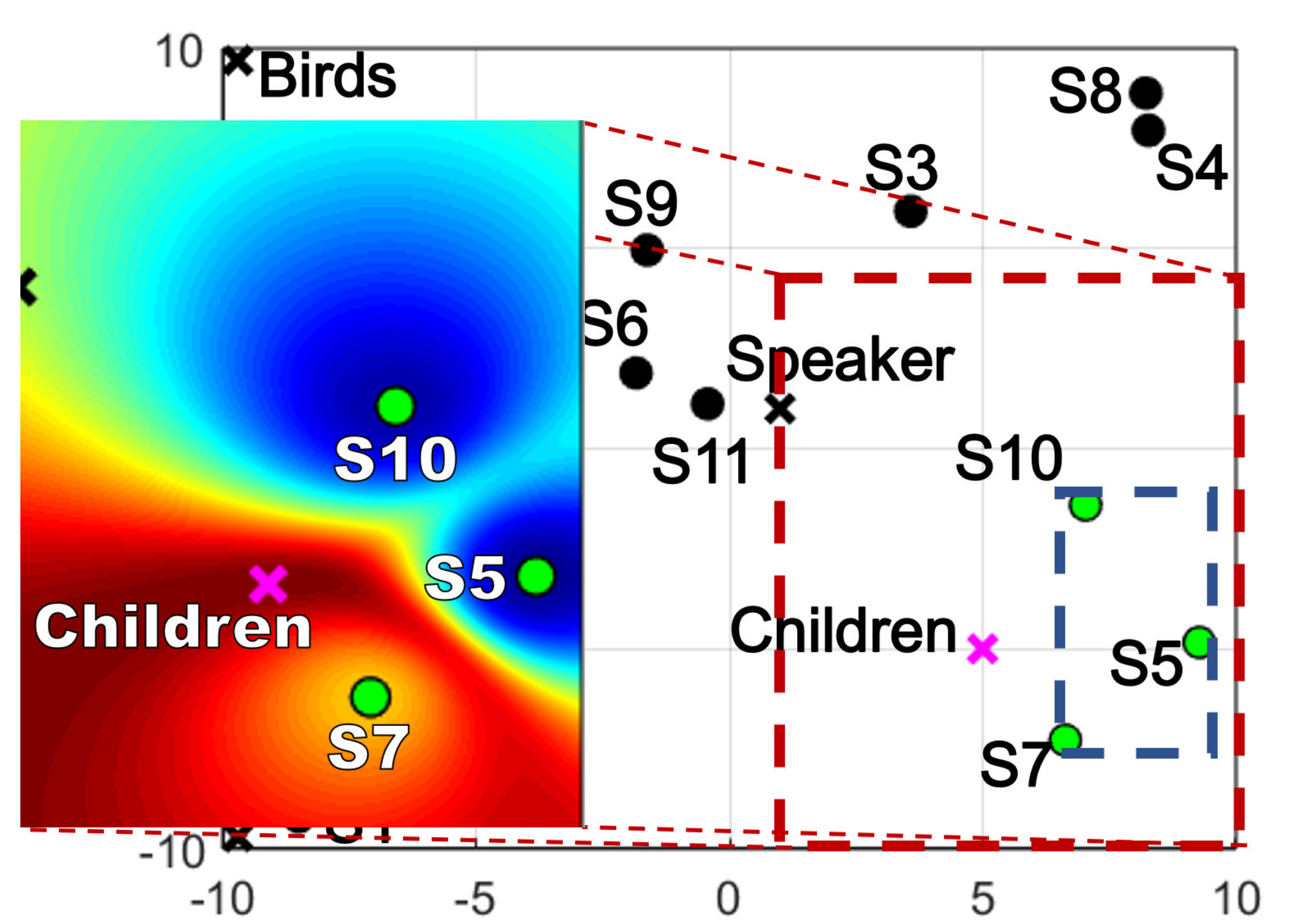}
         \vspace{-0.5cm}
         \caption{Scene:Park/Source:"children"}
         \end{subfigure}
         \begin{subfigure}[b]{0.24\textwidth}	         \includegraphics[width=\textwidth, height=3 cm]{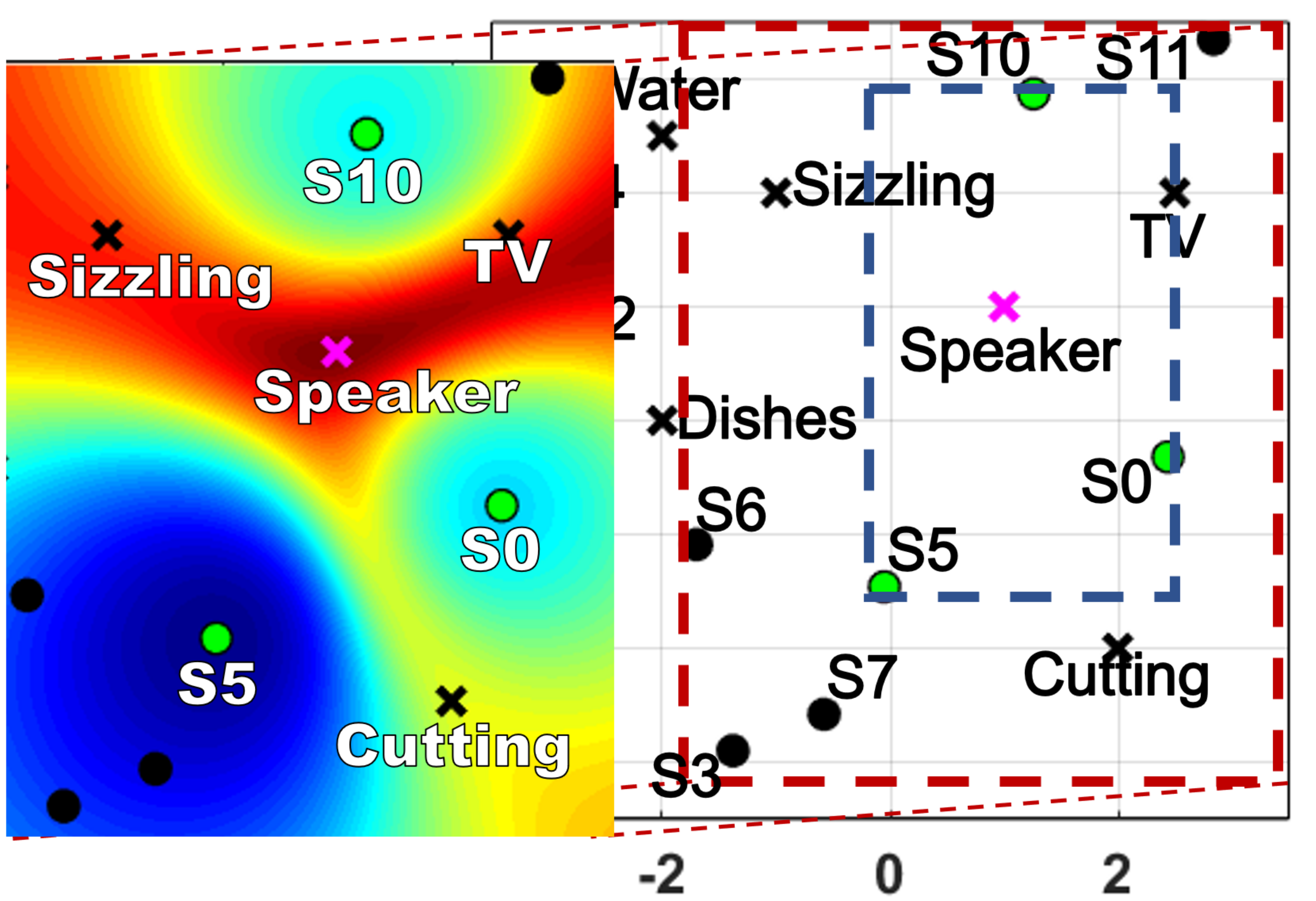}
         \vspace{-0.5cm}
         \caption{Scene:Kitchen/Source:"speaker"}
         \end{subfigure}
         \begin{subfigure}[b]{0.24\textwidth}	         \includegraphics[width=\textwidth, height=3 cm]{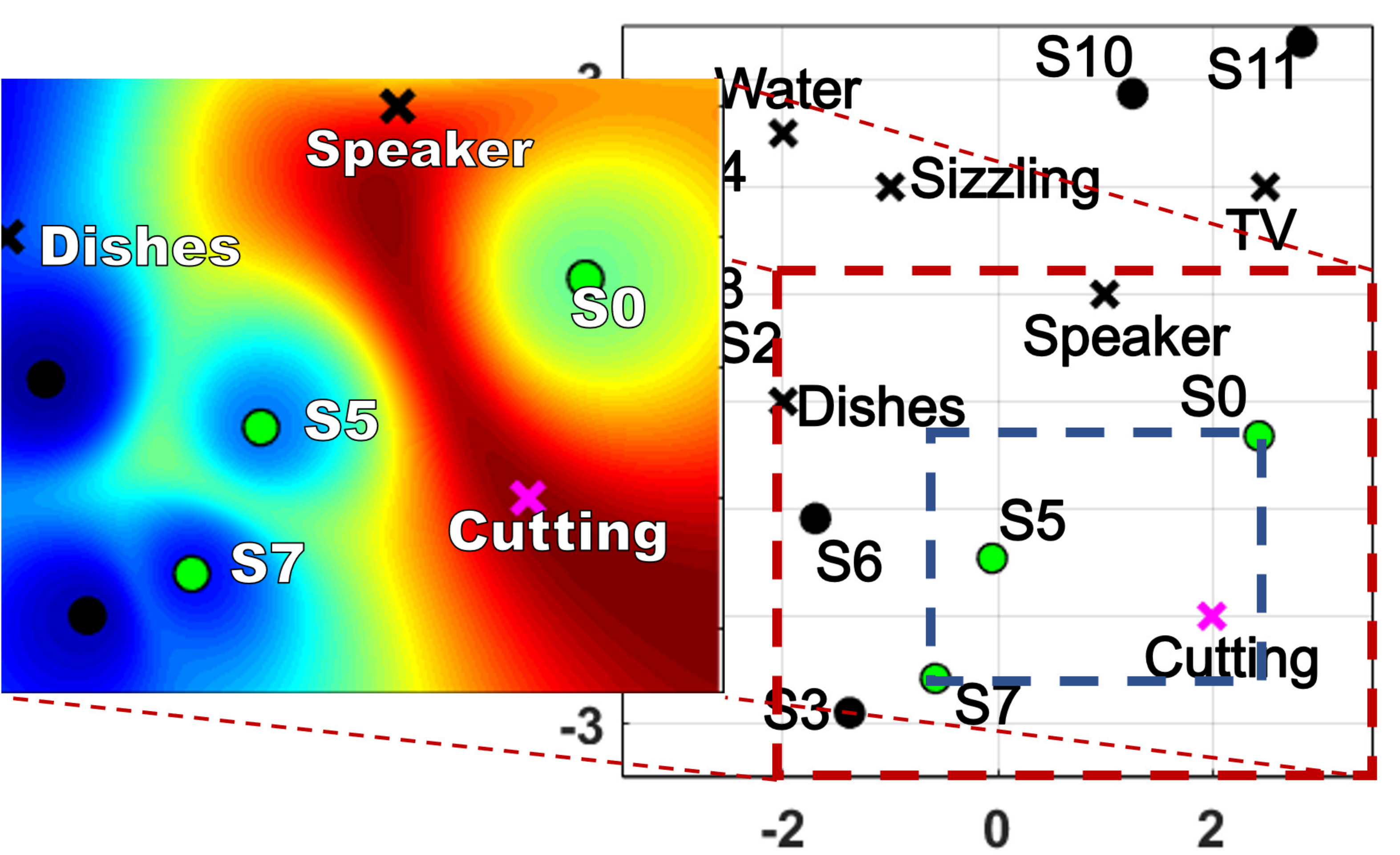}
         \vspace{-0.5cm}
         \caption{Scene:Kitchen/Source:"cutting"}
         \end{subfigure}
         \vspace{-0.25cm}
         \caption{ESFE method applied in different scenarios. The blue dashed rectangle is the smallest space delimited by the selected sensors and the red dashed rectangle is the new target source search area.}
        \vspace{-0.45cm}
        \label{fig:met}
	   \end{figure*}

Fig. \ref{fig:scene} illustrates acoustic energy distribution maps of Park and Kitchen scenes where the higher acoustic energy is represented in red and the lower in dark blue. The maps represents the simulation of the acoustic scenes that consists on the random distribution of target sources (magenta square), noise sources (black "$\times$") and microphones (green circles) in scene area. Energy distribution is estimated based on the sensor readings according to (7). Acoustic target and noise sources are non-stationary, then energy distribution estimation will vary at each signal frame. It also depends on the position and number of sensors distributed.


In this paper, the derivation of the Crámer-Rao Lower Bound (CRLB) is introduced in order to evaluate the performance of the H-ML-Energy estimator. The CRLB is a theoretical lower bound of the variance of an unbiased parameter estimate \cite{sheng2005maximum}\cite{djurovi2018achieving}. Firstly, the Fisher matrix $\mat{F}$ is calculated as $\mat{F}=-\mathbb{E}[\frac{\partial}{\partial\boldsymbol{\theta}}(\frac{\partial}{\partial\boldsymbol{\theta}}ln f(\mat{Z_H}|\boldsymbol{\theta})^T]$. From \eqref{eq:ml}, the Fisher matrix can be rewritten as $\mat{F}=\frac{\partial(\mat{DB})}{\partial\boldsymbol{\theta}}\mat{G_H^TG_H}\frac{\partial(\mat{DB})}{\partial\boldsymbol{\theta}^T}.$ The term $\frac{\partial(\mat{DB})}{\partial\boldsymbol{\theta}}$ can be derived as
\vspace{-0.3cm}
\begin{equation}
\vspace{-0.2cm}
\mat{C^T}_j=\frac{\partial(\mat{DB})^T}{\partial\rho_j}=-2B_j
\left[\frac{c_{1j}}{d^3_{1j}}  \cdots  \frac{c_{Nj}}{d^3_{Nj}}\right], 
\end{equation}where $c_{ij}=\frac{\partial{d_{ij}}}{\partial\rho_j}=\frac{(\rho_j-r_i)}{d_{ij}}$ is a unit vector from $j$-th source to the $i$-th sensor. Considering $\frac{\partial(\mat{DB})^T}{\partial\mat{B}}=\mat{D}$, then $\mat{F}$ can be expressed as
\vspace{-0.2cm}
\begin{equation}
\vspace{-0.2cm}
    \mat{F}=
    \begin{bmatrix}
    \mat{C}^T\\
    \mat{D}^T
    \end{bmatrix}\mat{G^TG}\begin{bmatrix}
    \mat{C} \mat{D}
    \end{bmatrix}.
\end{equation}Finally, the CRLB is computed as $\text{CRLB}=\sqrt{\sum^Q_{i=1}\frac{[\mat{F}^{-1}_{11}]+[\mat{F}^{-1}_{22}]}{Q}}$,where $Q$ is the number of blocks.

\vspace{-0.2cm}
\section{Proposed ESFE method}
\vspace{-0.25cm}

Acoustic sensing flow is defined as the area with the highest energy in the scene space. This flow is created by the energy emitted from the target sources. Hence, the WASN sensors placed near the source are mainly affected by the energy flow and thus forming an acoustic concentration region. The sensors signals in this region are less corrupted by the scene noise and consequently are more nonstationary. The ESFE consists on the selection of the sensors according to their $\text{INS}_\text{max}$ values that lead to new sensing flow area and thus search field reduction. The location estimation focuses on the most nonstationary sensor signals. For this purpose, let $x_i, i= 1, \ldots , L$, be the signals from the $L$ sensors. Selected signals are those with the highest values of $\text{INS}_\text{max}$. The selection criterion is defined as
\vspace{-0.2cm}
\begin{equation}
\vspace{-0.2cm}
\frac{|{\text{INS}_\text{max}}_i - max(\text{INS}_\text{max})|}{max(\text{INS}_\text{max})} \geq \alpha, 
\end{equation} 
where $max(\text{INS}_\text{max}) = \max_{1 \leq i \leq L} {\text{INS}_\text{max}}_i$, and the selection threshold $\alpha \in [0,1]$ is a function of the number of sensors ($L$) and the highest scene dimension ($v$) defined by the ESFE 
algorithm, i.e.,
\vspace{-0.2cm}
\begin{equation}
\vspace{-0.2cm}
\alpha=\frac{1}{\kappa(v+L)} + \xi,
\end{equation} where $\kappa$ is the measurements adjustment factor and $\xi$ is the estimation error. 
The values $\kappa=0.087$ and $\xi=\pm5\%$ were defined according to extensive experiments. Finally, source location is estimated using only the selected sensors. The vector $Z'_H$ of the normalized acoustic energy in the selected sensors is given by $\mat{Z_H'}= [\frac{u_1-\mu_{H_1}}{\sigma_{H_1}} \cdots \frac{u_N-\mu_{H_N}}{\sigma_{H_N}}]^T$.
The log-likelihood function is 
    $L(\boldsymbol{\theta})=||\mat{Z'_H}-\mat{G'_HD'B}||^2$,
where $N$ is the number of selected sensors, and $\mat{G_H}$ and $\mat{D_H}$ represent the gain and attenuation matrices of the selected sensors, respectively.  
The proposed scheme also leads to a reduction on the location search field based on the selected sensors positions. Fig. \ref{fig:met} shows the ESFE approach applied in four different scenarios. The green circles correspond to the selected sensors, and the black circles other sensors. The blue dashed rectangle represents the sensing flow area delimited by the selected sensors, and red one to the new target source search area. The new search area is the one delimited by the selected sensors with the addition of a security area that is defined according to the scene dimension. In these experiments, 20\% of $v$ were added to each side of the rectangle. The energy distribution in the energy flow cluster is also presented in the left of each scene. Note that the flow area has different sizes and location dependind on the scene and the target source.

For the evaluation of the proposed ESFE scheme, a method based on SNR is adopted for sensor selection. This approach was adopted in \cite{szurley2012energy} for large WASN ($L>80$) in order to save energy and extend network lifetime. This technique is here adapted to the energy based source localization methods and compared with the proposed ESFE. The sensors are chosen according to
the highest values of SNR \textit{a posteriori}, which is defined as $\text{SNR}_\text{post}=\frac{E[x(t)^2]}{E[n(t)^2]}=\frac{\sigma^2_x}{\sigma^2_n}$, where $x(t)$ is the noisy signal and $n(t)$ is the estimated noise given the time index $t$. According to the authors, the procedure stops when half of the sensors are selected by the algorithm.

\vspace{-0.25cm}
\section{Simulations and results}
\vspace{-0.25cm}

Extensive experiments are conducted to evaluate the accuracy improvement in energy based source localization methods. Networks of $L=12$ and $L=20$ omnidirectional sensors are randomly positioned in Park and Kitchen scenes. 
Each sequence has time duration of 3 seconds and is sampled at 16 kHz. 
One target source and five noise sources were chosen for each experiment.
The "speaker" is considered as the target source in both scenes while the other sources are assumed as noise. 
Furthermore, the source "children" in the Park scene and the source "cutting" in the Kitchen scene are adopted as target sources.

\begin{table}[t]
\vspace{-0.3cm}
\caption{$\text{INS}_\text{max}$ and $B_d$ results of each sensor.}
\vspace{-0.2cm}
\centering
\resizebox{0.5\textwidth}{!}{%
\begin{tabular}{|c||c|c||c|c||c|c||c|c|}\hline
&\multicolumn{4}{|c||} {Park} & \multicolumn{4}{|c|} {Kitchen}\\\cline{2-9}
Sensor&\multicolumn{2}{|c||} {"speaker" $\alpha=66\%$} & \multicolumn{2}{c||} { "children" $\alpha=63\%$}&\multicolumn{2}{|c||} {"speaker" $\alpha=28\%$} & \multicolumn{2}{c|} {"cutting" $\alpha=36\%$}  \\\cline{2-9}
&$\text{INS}_\text{max}$&$B_d$  &$\text{INS}_\text{max}$&$B_d$ &$\text{INS}_\text{max}$&$B_d$  &$\text{INS}_\text{max}$&$B_d$\\\hline \hline
S0	&	8.048	&	0.105	&	15.452	&	0.218	&	\textbf{43.696}	&	 \textbf{0.016}	&	\textbf{35.062}	&	 \textbf{0.010}	\\\hline
S1	&	3.072	&	0.091	&	7.694	&	0.157	&	21.442	&	0.065	&	12.503	&	0.067	\\\hline
S2	&	9.446	&	0.095	&	13.431	&	0.223	&	25.801	&	0.044	&	14.975	&	0.053	\\\hline
S3	&	 \textbf{18.955}	&	\textbf{0.081}	&	16.241	&	0.228	&	26.846	&	0.064	&	16.497	&	0.045	\\\hline
S4	&	8.157	&	0.116	&	14.478	&	0.243	&	20.708	&	0.049	&	12.058	&	0.061	\\\hline
S5	&	9.000	&	0.113	&	\textbf{30.379}	&	\textbf{0.132}	&	\textbf{36.499}	&	\textbf{0.030}	&	\textbf{27.527}	&	\textbf{0.017} 	\\\hline
S6	&	\textbf{32.921}	&	\textbf{0.039}	&	16.572	&	0.204	&	32.535	&	0.039	&	17.636	&	0.041	\\\hline
S7	&	11.077	&	0.110	&	\textbf{44.143}	&	\textbf{0.079}	&	27.331	&	0.053	&	\textbf{22.717}	&	\textbf{0.030} 	\\\hline
S8	&	5.962	&	0.118	&	13.048	&	0.246	&	21.722	&	0.053	&	12.370	&	0.065	\\\hline
S9	&	\textbf{20.707}	&	\textbf{0.070}	&	14.485	&	0.220	&	24.387	&	0.049	&	14.952	&	0.054	\\\hline
S10	&	16.104	&	0.091	&	\textbf{30.661}	&	\textbf{0.127}	&	\textbf{43.525}	&	\textbf{0.015}	&	12.060	&	0.056	\\\hline
S11	&	\textbf{55.571}	&	\textbf{0.005}	&	17.910	&	0.162	&	28.922	&	0.038	&	11.171	&	0.066	\\\hline
\end{tabular}}
\label{tab:ins}
\vspace{-0.3cm}
\end{table}

Table \ref{tab:ins} shows the $\text{INS}_\text{max}$ values for $x_i$ on four different scenarios 
considering 12 sensors network and the selection threshold $\alpha$. 
The Bhattacharrya distance ($B_d$) \cite{kailath1967divergence} is here used to confirm the efficiency of $\text{INS}_\text{max}$ as a sensor selection parameter. 
$B_d$ compares two probability distributions, $p_1(x)$ and $p_2(x)$, of two acoustic signals, $s_1(t)$ and $s_2(t)$, 
as $B_d=-\ln\int\sqrt{p_1(x)p_2(x)}dx$. In this paper $s_1(t)$  corresponds to the original target source signal and $s_2(t)$ to the sensor signals at each scenario. The sensors signals less corrupted by the scene are going to have the lowest values of $B_d$. Note that, as expected, the sensor with lower $B_d$ corresponds to the ones with higher $\text{INS}_\text{max}$ proving that the INS can be used as selection parameter. The selected sensors are highlighted in bold numbers.

\begin{figure}[!t]
\vspace{-0.65cm}
\centering
\includegraphics[width=0.95\linewidth]{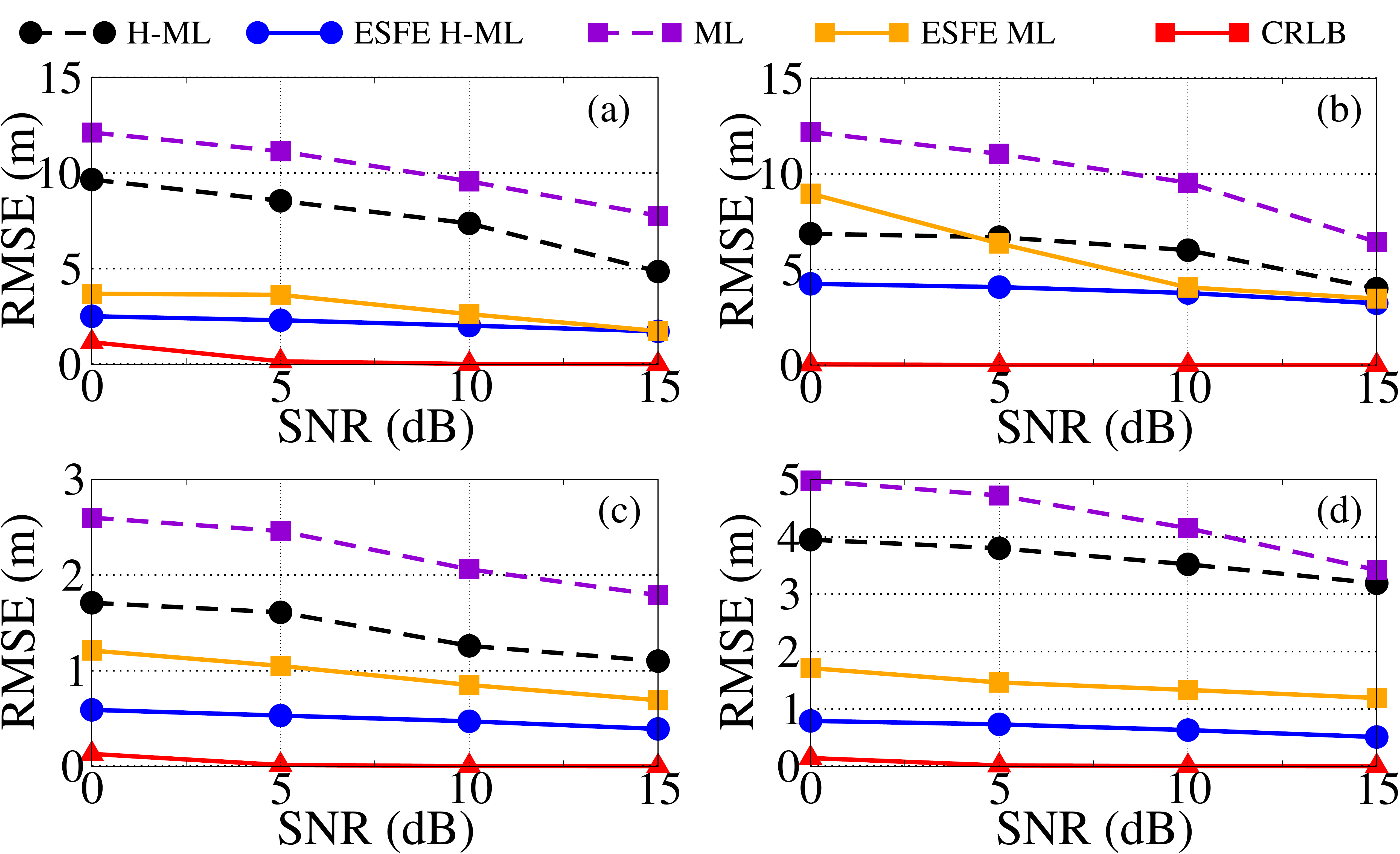}
\vspace{-0.25cm}
\centering
\caption{RMSE analysis L=12 sensors. (a) Park /"speaker",\\(b) Park /"children",(c) Kitchen /"speaker",(d) Kitchen /"cutting".}
\vspace{-0.4cm}
\label{fig:rmse12}
\end{figure}

The source localization estimation is conducted using energy-based methods, H-ML-Energy and ML-Energy, before and after the application of the ESFE scheme. 
Four different SNR conditions are evaluated: from 0 dB to 15 dB, with 5 dB increments. 
The SNR is calculated in a position 1 meter away from the source position. 
Blocks of $M$=1024 samples, i.e., a total of 46 blocks, are used in the evaluation experiments. 
Therefore, for each scene and each WASN with 12 and 20 sensors deployment, 552 tests are conducted
considering 3 different target sources ("speaker", "children", and "cutting") and 4 SNR values.
The gain of the sensors are set to $g_i=1$. The minimum of the log-likelihood function of the H-ML-Energy and the ML-Energy are found using the multiresolution search \cite{sheng2005maximum} with 0.1 meter in the Park scene (20 m x 20 m), and 0.035 meter in the Kitchen scene (7 m x 7 m). The root mean squared error (RMSE) is applied as evaluation measure in the experiments. It is defined as $\text{RMSE}=\sqrt{\frac{1}{Q}\sum^Q_{i=1}|\hat{r_i}-r_i|^2}$, where $r_i$ denotes the target source location of $i$-th $(i=1,2,...,Q)$ block and $\hat{r_i}$ represents its estimated position. The RMSE is used to verify how close the estimated localization are from the target source positions.

\begin{figure}[!t]
\vspace{-0.15cm}
\centering
\includegraphics[width=0.95\linewidth]{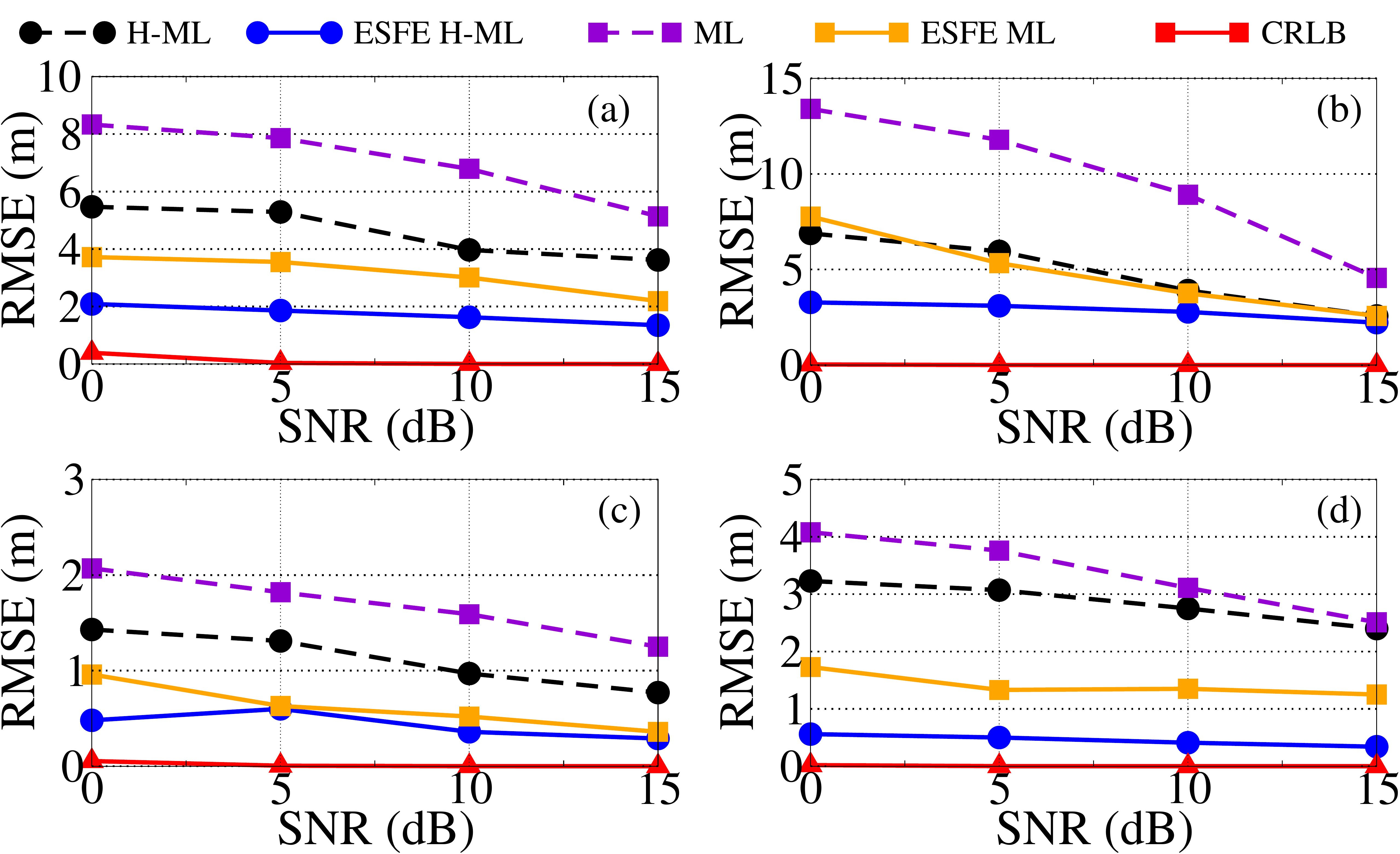}
\vspace{-0.2cm}
\caption{RMSE analysis L=20 sensors. (a) Park /"speaker",\\(b) Park /"children",(c) Kitchen /"speaker",(d) Kitchen /"cutting".}
\vspace{-0.2cm}
\label{fig:rmse20}
\end{figure}

Figs. \ref{fig:rmse12} and \ref{fig:rmse20} depict the RMSE values obtained with 12 and 20 sensors, respectively. 
They include H-ML-Energy and ML-Energy, the proposed ESFE for the H-ML-Energy and ML-Energy, and the CRLB estimation. 
Note that the application of the proposed scheme in the energy based location methods reduced the error estimation for studied scenarios and made it approaches the CRLB. 
The new scheme was more effective for the source "children" in the Park scene ($L=12$), RMSE values in the ML reduced from 12.19m to 2.66m in 0dB. The lowest reduction was for the source "speaker" in the Kitchen scene ($L=20$) where the RMSE varied from 0.77m to 0.29m in 15dB. It can also be observed that the H-ML-Energy outperforms the ML-Energy before and after the ESFE application in both scenarios, mainly for low SNR. This is explained by the fact that the H-ML-Energy takes in consideration the correlation between the signal from the source and the interference of the scene.

\begin{table}[!t]
\vspace{-0.2cm}
    \caption{Normalized mean processing time.}
\vspace{-0.2cm}
    \centering
    \begin{tabular}{|c|c||c|c|}\hline
         H-ML-12  & ESFE-12 & H-ML-20  & ESFE-20  \\ \hline\hline
         3.33     & 1.00    & 4.30     & 1.19\\ \hline
    \end{tabular}
    \vspace{-0.3cm}
    \label{tab:complex}
\end{table}

Table \ref{tab:complex} indicates the computational complexity which refers to the processing time required for each algorithm evaluated for 1024 samples per frame. In order to compare to the processing time of the different methods, the values were normalized according to the ESFE method with $L=12$ sensors as reference (execution time = 1). Note the new approach not only improves the localization estimation accuracy, but also decreases the processing period more than 3 times in both network sizes.

\begin{figure}[t]
\vspace{-0.65cm}
\centering
\includegraphics[width=\linewidth]{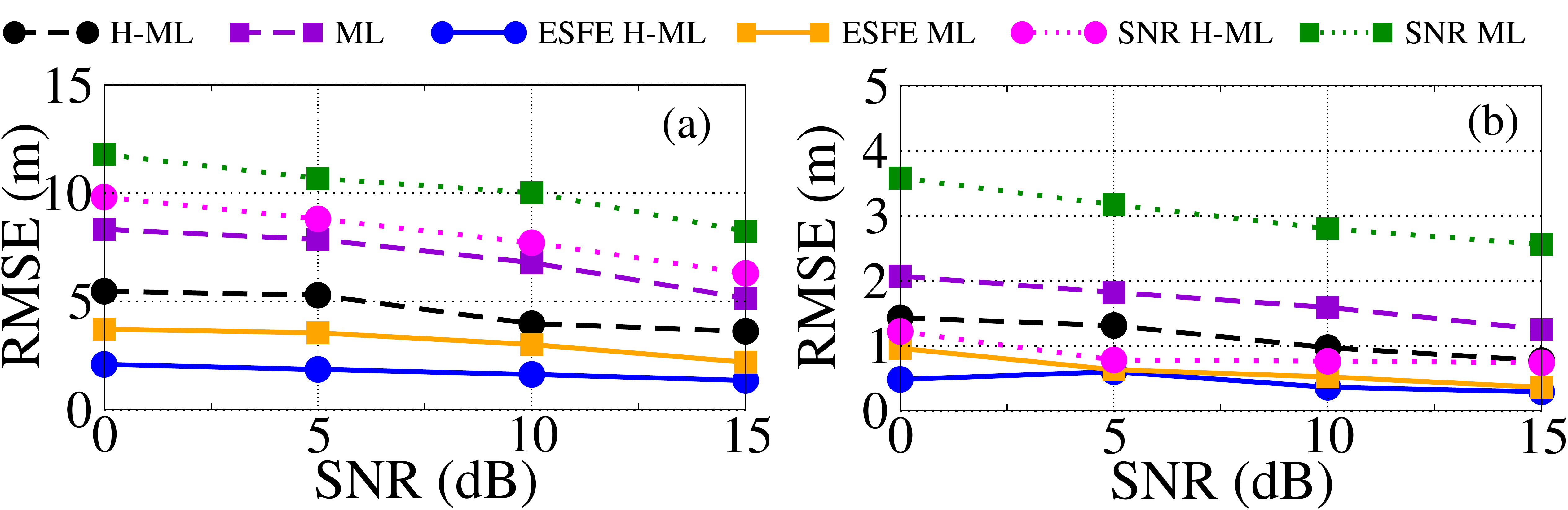}
\vspace{-0.5cm}
\centering
\caption{RMSE comparison analyses.(a)Park, (b)Kitchen.}
\vspace{-0.3cm}
\label{fig:snr}
\end{figure}

The selection of sensors based on SNR was evaluated and compared with the proposed method for the Park and Kitchen scenes with $L=20$. The target source was the "speaker", since SNR-based selection methods depend on voice activity detectors (VAD) \cite{szurley2012energy}. Fig. \ref{fig:snr} presents the RMSE values for the different approaches. 
The best results of the SNR based selection were achieved in the Kitchen scene for H-ML-Energy method.

\vspace{-0.2cm}
\section{Conclusion}
\vspace{-0.2cm}

This letter introduced an effective acoustic energy sensing approach, ESFE, to improve the accuracy of energy based source localization methods in acoustic scenes. The new scheme detects the scene energy flow based on the nonstationarity index of the sensor signals readings. Several experiments were conducted with different acoustic scenes and target sources. The results demonstrated that the proposed approach consistently improves the localization estimation while reducing the number of sensors.

\vspace{-0.3cm}

\section*{Acknowledgment}
\vspace{-0.2cm}
\addcontentsline{toc}{section}{Acknowledgment}
\scriptsize
R. Coelho work was supported in part by the National Council for Scientific and Technological Development (CNPq) and Fundação de Amparo à Pesquisa do Estado do Rio de Janeiro (FAPERJ) under 307866/2015 and  203075/2016 research grants. The work was also supported by the Coordenação de Aperfeiçoamento de Pessoal de Nível Superior (CAPES) - Grant Code 001.

\normalsize

\vspace{-0.2cm}

\bibliographystyle{IEEEtran}
\bibliography{ref.bib}
\vspace{-0.3cm}

\end{document}